\documentclass[11pt,a4paper]{article}
\usepackage{amsmath,amssymb}
\usepackage{epsfig,graphicx}

\newcommand{\vecc}[1]{\mbox{\boldmath $#1$}}
\def\ex{\hbox{e}}

\def\e{\epsilon}

\def\pd{\partial}

\def\F{\Phi}

\def\pb{\bar \psi}

\def\z{\zeta}

\def\ex{\hbox{e}}
\def\S{\Sigma}
\def\F{\Phi}
\def\<{\langle}
\def\>{\rangle}

\def\a{\alpha}

\def\d{\delta}

\def\m{\mu}
\def\n{\nu}

\def\z{\zeta}

\def\({\left(}
\def\[{\left[}
\def\){\right)}
\def\]{\right]}

\def\pd{\partial}

\def\pa{{\cal P}}

\def\halb{\frac{1}{2}}

\begin{document}

\title{{\hfill {\footnotesize RUB-TPII-08/08}}\\ [1cm]\bf New results on gauge-invariant
       TMD PDFs in QCD\thanks{Talk presented by the first author at QUARKS-2008
       15th International Seminar on High Energy Physics, Sergiev Posad, Russia,
       23-29 May, 2008}}
\author{I.~O.~Cherednikov$^a$\footnote{{\bf E-mail}: igor.cherednikov@jinr.ru},
N.~G.~Stefanis$^{b}$\footnote{{\bf E-mail}: stefanis@tp2.ruhr-uni-bochum.de}
\\
$^a$ \small{\em Bogoliubov Laboratory of Theoretical Physics, JINR} \\
\small{\em RU-141980 Dubna, Russia}\\
$^b$ \small{\em Institut f\"{u}r Theoretische Physik II,
             Ruhr-Universit\"{a}t Bochum} \\
\small{\em D-44780 Bochum, Germany}
}
\date{}
\maketitle

\begin{abstract}
The renormalization properties of unintegrated
(trans\-ver\-se\--mo\-men\-tum dependent) parton distribution
functions (TMD PDF's) are used for analyzing their completely
gauge-invariant definition.
To this end, the UV anomalous dimension is calculated at
the one-loop order in the light-cone gauge and a consistent
treatment of the extra singularities, which produce undesirable
contributions in the anomalous dimensions, is given.
The generalized definition of a TMD PDF, based
on the renormalization procedure for the Wilson exponentials with
obstructions, is proposed.
The reduction of the re-defined TMD PDF to the integrated PDF's, as
well as their probabilistic interpretation, are discussed.
\end{abstract}

\paragraph{Introduction}
Parton distribution functions (PDF's) play a crucial role in QCD
phenomenology \cite{Col03, BR05, Col08}.
In inclusive processes (e.g., DIS), the standard (integrated)
PDF's, which originate from the parton model, are used.
The integrated PDF's depend on the longitudinal fraction of the
momentum, $x$, and on the scale of the hard subprocess $Q^2$.
The completely gauge invariant definition of integrated PDF's reads
\begin{equation}
  \hat f_{i}(x)
=
  \frac{1}{2} \int \frac{d\xi^-}{2\pi}\
  \ex^{- i k^+ \xi^-} { \langle h(P) | } \bar \psi_i
  (\xi^-, \vecc 0_\perp) [\xi^-, 0^-]\gamma^+
  \psi_i(0^-,\vecc 0_\perp) { |h(P)\rangle } \ ,
\end{equation}
where the Wilson line (gauge link), ensuring gauge invariance, is
defined as follows
\begin{equation}
  { [y,x|\Gamma] }
=
  {\cal P} \exp
  \left[-ig\int_{x[\Gamma]}^{y}dz_{\mu} A_{a}^{\mu}(z) t_{a}
  \right] \ .
\end{equation}
The renormalization properties of these objects are described
by the DGLAP evolution equation
\begin{equation}
  \mu \frac{d}{d\mu} f_{i} (x, \mu)
=
  \sum_j \int_x^1\! \frac{dz}{z} \ P_{ij}
  \left(\frac{x}{z}\right) f_{j} (z, \mu) \ ,
  \label{eq:dglap}
\end{equation}
where $P_{ij}$ is the DGLAP integral kernel.
The renormalization properties of the quantities under consideration
(to be precise, their anomalous dimensions)
are the cornerstone of our approach.
The reason is that anomalous dimensions (within perturbative QCD)
accumulate the main characteristics of Wilson lines in \textit{local}
form, while the gauge contours are \textit{global} objects and,
therefore, complicated to handle within a local-field theory framework.

\paragraph{Unintegrated PDF's}

The study of semi-inclusive processes, such as SIDIS,
or the Drell-Yan process, where the transverse momentum of
the produced hadrons can be observed,
requires the introduction of more complicated quantities, so-called
unintegrated, or transverse-momentum dependent, PDF's.
In this case, one does not integrate over the transverse
component of the parton's momentum
$\vecc k_\perp$, and the corresponding distribution function
looks like a generalization of the integrated PDF.
The ``naive'' definition we start with reads
\begin{eqnarray}
  f_{i} (x, \vecc k_\perp)
\!\!\!\! && \!\!\!\! =
  \frac{1}{2}
  \int \frac{d\xi^- d^2\xi_\perp}{2\pi (2\pi)^2} \
  \ex^{
       - i k^+ \xi^-
       + i \mbox{\footnotesize \boldmath$k_\perp$} \cdot
           \mbox{\footnotesize \boldmath$\xi_\perp$}
       }
  {\< p | } \pb_i (\xi^-, \xi_\perp)
  [\xi^-, \vecc \xi_\perp; \infty^-, \vecc \xi_\perp;]^\dagger
\nonumber \\
&& \times
  \gamma^+
   [\infty^-, \vecc 0_\perp; 0^-, \vecc 0_\perp]
   \psi_i (0^-,0_\perp) { | p \> } \ \ .
\end{eqnarray}
Formally, the integration over the transverse component of the parton's
momentum is expected to yield the integrated distribution
\begin{equation}
  \int\! d^2 k_\perp f_i (x, \vecc k_\perp)
=
  \hat f_{i/h} (x)\ .
\end{equation}

However, this definition, taken literally, suffers from several
shortcomings
(see, e.g, the recent works in Refs.\ \cite{Col08,CRS07,Bacch08,CS07}):
\begin{itemize}
\item Gauge invariance, in fact, is not complete: in the light-cone
gauge, the dependence on the pole prescription in the gluon
propagator remains.

\item Extra (rapidity) divergences arise, which are associated with
the known features of the light-cone gauge, or the light-like Wilson
lines, that cannot be removed by ordinary ultraviolet renormalization
alone.
Note, that in the integrated case, these divergences, though
appearing at the intermediate steps of the calculation,
they are absent in the final result due to the mutual
calcellation between real and virtual gluon contributions.

\item The reduction to the integrated case cannot be performed
straightforwardly: the formal integration does not
reproduce the correct result (i.e., the DGLAP kernel)
because of additional uncanceled UV divergences.
\end{itemize}

The following methods to take care of the above-mentioned
problems have been proposed in the literature:
\begin{itemize}
\item Gauge invariance is restored by means of  {an} additional
transverse Wilson line at light-cone infinity
\cite{JY02, BJY03, BMP03}.
This gauge link contributes only in the light-cone gauge and
cancels the pole-prescription dependence.

\item Extra divergences can be avoided by using the non-light-like
gauge connectors in covariant gauges, or an axial
gauge off the light cone \cite{CS81, JMY04}.
This, however, entails the introduction of an additional rapidity
parameter
$\zeta=(p\cdot n)^{2}/n^{2}$ (with $n^{2}\neq 0$)
to encode the deviation from the light cone.
To establish the independence from this arbitrary variable, an
additional evolution equation to the standard one has to be
fulfilled rendering the reduction to the integrated case
questionable.
Besides, factorization off the light cone also becomes problematic.

\item Application of a generalized renormalization procedure
for the light-like Wilson lines (or a subtractive method):
as a result, extra divergences cancel by the additional ``soft''
factor, defined by the vacuum average of particular Wilson lines
(demonstrated explicitly in a covariant gauge, in the 1-loop
order) in \cite{CH00, Hau07}, see also \cite{CM04}.

\end{itemize}

In this work, we implement the analysis of anomalous dimensions
within the latter approach.
This allows us to figure out the necessary modifications of
TMD PDF's in the most economic way.
Towards a ``completely correct'' definition, we calculate
the anomalous dimension of the TMD PDF (in fact, we calculate
the distribution of a ``quark in a quark'') in the light-cone
gauge and identify extra UV divergences in terms of the
entailed defect of the anomalous dimension.
Then, we  perform a generalized renormalization procedure
of the TMD PDF, similar to the renormalization of
the Wilson contours with cusps or self-intersections
\cite{Pol80, CD80}.
This renormalization cancels undesirable divergences and yields
a completely gauge invariant definition of TMD PDF's.

\paragraph{One-loop anomalous dimension}

\begin{figure}
\centering
\vspace*{0mm} \scalebox{1.3}{\includegraphics[width=0.3\textwidth,angle=90]{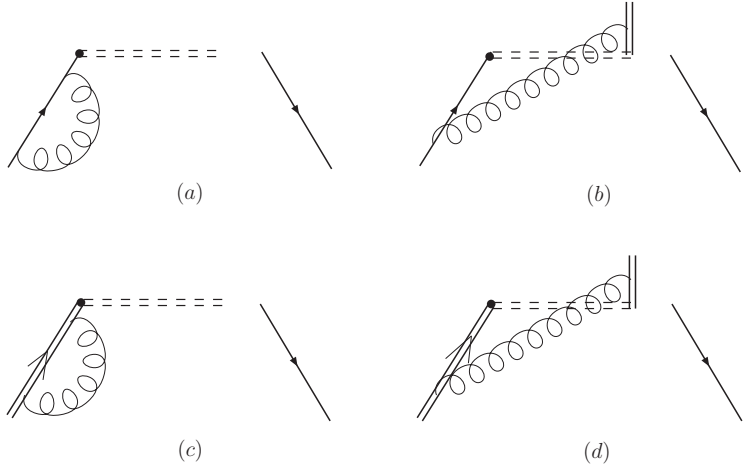}}
 \caption{
One-gluon exchanges  {in} the TMD PDF in the light-cone gauge.
Diagrams $(a)$ and $(b)$ produce UV divergences in the considered TMD
PDF, while diagrams $(c)$ and $(d)$ correspond to the UV divergences of
the soft factor.
The conjugated (``mirror'') diagrams are not shown.}
\label{fig:1}
\end{figure}

In the tree approximation, the TMD PDF reads
\begin{equation}
  f^{(0)} (x, \vecc k_\perp)
=
  \d(1 -x ) \d^{(2)} (\vecc k_\perp) \ .
\end{equation}
The one-gluon exchanges, contributing to the UV-divergences,
are described by the diagrams Fig.\ \ref{fig:1}$(a, b)$.

The source of the uncertainties and extra divergences is the pole in
the gluon propagator in the light-cone gauge:
\begin{equation}
  D^{\m\n}_{\rm LC} (q)
= \frac{-i}{q^2}
  \[g^{\m\n} - {\frac{q^\m n^{-\n}}{[q^+]}
             - \frac{q^\n n^{-\m} }{[q^+]}}
  \] \ ,
\end{equation}
where $[q^+]$ stands for an undefined denominator.
Consider now the following pole prescriptions:
\begin{equation}
  \frac{1}{[q^+]}_{\rm PV}
=
  \halb \( \frac{1}{q^+ + i \eta} + \frac{1}{q^+ - i \eta} \) \ \
\hbox{and} \ \
  \frac{1}{[q^+]}_{\rm Adv/Ret}
=
  \frac{1}{q^+ \mp i \eta}  \ .
\end{equation}
In what follows, we keep $\eta$ small, but finite.
To control UV singularities, dimensional regularization is used.
Another widely used prescription is the Mandelstam-Leibbrand
(ML) one:
\begin{equation}
  \frac{1}{[q^+]}_{\rm ML}
  =
  \frac{1}{q^+ + i \eta q^-} =  \frac{q^-}{q^+ q^- + i \eta }
\end{equation}
to be considered in a separate work.

The UV divergent part of the diagrams 1($a,b$) (without their
``mirror'' contributions) reads
\begin{equation}
  {\S}^{UV}_{\rm left} (p, \a_s ; \e)
=
  {
   - \frac{\a_s}{\pi}C_{\rm F} \ \frac{1}{\e}
   \[- \frac{3}{4} - \ln \frac{\eta}{p^+} + \frac{i\pi}{2}
    + i \pi \ C_\infty \] + \a_s C_{\rm F} \ \ \frac{1}{\e}
    \[i C_\infty\]
  } \ ,
\end{equation}
where $C_{\rm F}=(N_{\rm c}^{2}-1)/2N_{\rm c}=4/3$ and the numerical
factor $C_{\infty}$ accumulates the pole-prescription uncertainty,
being defined by
\begin{equation}
C_\infty
   =
   \left\{
   \begin{array}{ll}
   & \  0  \ , \ {  {\rm Adv:} \ \frac{1}{[q^+]}
   = \frac{1}{q^+-i\eta} } \   \\
   & - 1 \ , \ {  {\rm Ret:} \ \frac{1}{[q^+]}
   = \frac{1}{q^++i\eta}  } \ \\
   & - \frac{1}{2} \ , \ {   {\rm PV:} \ \ \frac{1}{[q^+]}
   = \frac{1}{2}\(\frac{1}{q^+-i\eta} + \frac{1}{q^++i\eta} \) }
   \
   \end{array} \right.  .
\end{equation}
One immediately observes that the prescription dependence is canceled
due to the contribution of the transverse gauge link at the
light-cone infinity---diagram 1($b$).
Taking into account the ``mirror'' contributions (designated as
``right'' below), one gets the total real UV divergent part:
\begin{equation}
  \S_{\rm tot}^{UV} (p, \a_s (\m) ; \e )
=
  \S_{\rm left} +  \S_{\rm right} =
  {  - \frac{ \a_s}{4\pi}C_{\rm F} \   \frac{2}{\e}
  \(- 3 - 4 \ln \frac{\eta}{p^+} \) } \ .
\label{eq:tot_uv}
\end{equation}

The one-loop anomalous dimension is defined via the renormalization
factor
\begin{equation}
  \gamma
=
  \halb \ \frac{1}{Z^{(1)}} \ \m \frac{\pd \a_s (\m)}{\pd\m}
  \ \frac{\pd Z^{(1)} (\m, \a_s (\m); \e)} {\pd\a_s}
\end{equation}
and, using Eq.\ (\ref{eq:tot_uv}), it reads
\begin{equation}
  \gamma_{\rm LC}
=
  \gamma_{\rm smooth} - {  \d \gamma } \ \ , \ \ \gamma_{\rm smooth}
=
  {  \frac{3}{4} \frac{\a_s}{\pi}C_{\rm F} } + O(\a_s^2) \ .
\end{equation}
Here we introduce the defect of the anomalous dimension
\begin{equation}
  \d \gamma
=
  - \frac{ \a_s}{\pi}C_{\rm F} \  \ln \frac{\eta}{p^+} \ ,
\end{equation}
which marks the deviation of the calculated quantity from the
anomalous dimension of the two-quark operator with the smooth
(i.e., direct) gauge connector.
The latter equals the double anomalous dimension of the fermion field,
while $\gamma_{\rm LC}$ contains an undesirable $p^+$-dependent
term that should be removed by an appropriate procedure.
Note that $p^+ = (p \cdot n^-) \sim \cosh \chi$
defines, in fact, an angle $\chi$ between the direction of the quark
momentum $p_\mu$ and the light-like vector $n^-$.
In the large $\chi$ limit,
$\ln p^+ \to \chi , \ \chi \to \infty$.
Thus, we can conclude that the defect of the anomalous dimension,
$\delta \gamma$, can be identified with the well-known cusp anomalous
dimension \cite{KR87}.

\paragraph{Generalized renormalization}

\begin{figure}
\centering
\vspace*{0mm} \scalebox{.8}{\includegraphics[width=0.9 \textwidth,angle=0]{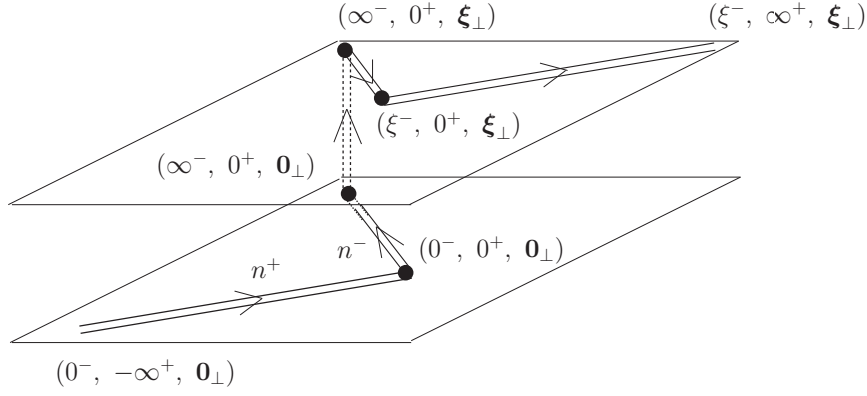}}
 \caption{
Integration  contour for the additional cusp-dependent
renormalization factor.}
\label{fig:2}
\end{figure}

It is known that the renormalization of the Wilson operators
with obstructions (cusps, or self-intersections) cannot be performed
by the ordinary $R-$operation alone, but requires an
additional renormalization factor depending on the cusp angle
\cite{Pol80,CD80,Aoy81,KR87}:
\begin{equation}
Z_{p^+} = \[\left\<0 \left| \pa \exp\[ig \int d\z^\m
   \ \hat A^a_\m (\z)\] \right|0\right\> \]^{-1} \ .
\end{equation}
Using this statement as a hint, we compute the extra renormalization
constant associated with the soft counter term \cite{CH00} and show that
it can be expressed in terms of a vacuum expectation value of a
specific gauge link.
Hence, in order to cancel the anomalous dimension defect
$\delta \gamma$,
we introduce the counter term
\begin{equation}
  R
\equiv
 \Phi (p^+, n^- | 0) \Phi^\dagger (p^+, n^- | \xi) \ ,
\label{eq:soft_factor_1}
\end{equation}
where
\begin{equation}
  \Phi (p^+, n^- | \xi )
 =
  \left\langle 0
  \left| {\cal P} \exp\Big[ig \int_{\Gamma_{\rm cusp}}d\zeta^\mu
  \ t^a A^a_\mu (\xi + \zeta)\Big]
  \right|0
  \right\rangle
\label{eq:soft_definition}
\end{equation}
and evaluate it along the non-smooth, off-the-light-cone integration
contour $\Gamma_{\rm cusp}$, depicted in Fig.\ \ref{fig:2}.

The one-loop gluon virtual corrections, contributing to the UV
divergences of the soft factor $R$, are shown in Fig.\
\ref{fig:1}($c,d$).
For the UV divergent term we obtain
\begin{equation}
  \Sigma_{R}^{\rm UV}
=
  - \frac{ \alpha_s}{\pi} C_{\rm F} \ 2 \left(  \frac{1}{\epsilon} \
  \ln \frac{\eta}{p^+} - \gamma_E + \ln 4 \pi \right)
\end{equation}
and observe that this expression is equal, but with opposite sign,
to the unwanted term in the UV singularity, related to the cusped
contour, calculated above.

Therefore, we propose to  redefine the conventional TMD PDF and
absorb the soft counter term in its definition:
$$
   f_{q/q}^{\rm mod}(x, \mbox{\boldmath$k_\perp$})
 =
  \frac{1}{2}
   \int \frac{d\xi^- d^2
   \vecc \xi_\perp}{2\pi (2\pi)^2}
   {\rm e}^{- i k^+ \xi^- + i 
   {\vecc k_\perp} \cdot {\vecc \xi_\perp}}
   \bigl\langle  q(p) |\bar \psi (\xi^-, \xi_\perp)
   [\xi^-, \mbox{\boldmath$\xi_\perp$};
   \infty^-, \mbox{\boldmath$\xi_\perp$}]^\dagger
   [\infty^-, \mbox{\boldmath$\xi_\perp$};
   \infty^-, \mbox{\boldmath$\infty_\perp$}]^\dagger 
$$
\begin{equation}
   \times \gamma^+[\infty^-, \mbox{\boldmath$\infty_\perp$};
   \infty^-, \mbox{\boldmath$0_\perp$}]
   [\infty^-, \mbox{\boldmath$0_\perp$}; 0^-,\mbox{\boldmath$0_\perp$}]
   \psi (0^-,\mbox{\boldmath$0_\perp$}) |q(p)\bigr\rangle
   \cdot
   R (p^+, n^-) \ .
\label{eq:tmd_re-definition}
\end{equation}

One immediately verifies that the integration over the transverse
momentum $\vecc k_\perp$ yields the integrated PDF:
\begin{equation}
   \int\! d^{\omega-2} \vecc k_\perp
   f_{i/a}^{\rm mod} (x, \vecc k_\perp ; \mu , \eta)
   =
   f_{i/a} (x, \mu) \ ,
\end{equation}
which obeys the DGLAP equation (\ref{eq:dglap}).
The anomalous dimension of the modified TMD PDF
(\ref{eq:tmd_re-definition}) is equal to the anomalous dimension of
the corresponding operator with the smooth gauge connector, according
to the anomalous dimensions (AD) sum rule, which can be formulated in
the following symbolic form
$$
  {\rm  AD} \  \halb \int \frac{d\xi^- d^2\xi_\perp}{2\pi (2\pi)^2}
   \ex^{- i k^+ \xi^-
   + i \mbox{\footnotesize \boldmath$k_\perp$}
   \cdot
       \mbox{\footnotesize \boldmath$\xi_\perp$}} \
   { \< p|}\bar \psi (\xi)\gamma^+ {\[\xi, 0\]_{\rm direct \ link} }
   \psi (0) {  |p \> }
=
$$
\begin{equation}
   {\rm  AD} \ \halb \int \frac{d\xi^- d^2\xi_\perp}{2\pi (2\pi)^2}
   \ex^{- i k^+ \xi^-
   + i \mbox{\footnotesize \boldmath$k_\perp$}
   \cdot
       \mbox{\footnotesize \boldmath$\xi_\perp$}}
  { \< p| }\bar \Psi (\xi | \infty)\gamma^+ \Psi (0 | \infty) { |p \> }
  { \F(p^+, n^- | 0^-, \vecc 0_\perp)
  \F^\dag (p^+, n^- | \xi^-, \vecc \xi_\perp) } \ .
\end{equation}

This sum rule is based on the following considerations, based on the
probabilistic interpretation of PDF's in terms of their anomalous
dimensions (alias the RG properties) of corresponding operators.
The distribution functions cannot be calculated
from first principles, but their evolution can.
In particular, we have the DGLAP equation for the integrated
PDF's and the two-quark UV anomalous dimension for the
TMD PDF's, where the quark fields are separated by a non-light-like
distance.
The requirement that the off-the-light-cone two-quark matrix
element should have an anomalous dimension equal to that of
the corresponding quantity with the smooth gauge connector in order to
respect the probabilistic interpretation, is tantamount to the
anomalous dimensions sum rule.
Therefore, the RG properties can serve to define the necessary
condition for the PDF to be a number density.
The generalized TMD PDF indeed obeys this condition.

It is interesting to note that the additional soft counter term $R$ can
be treated within Mandelstam's explicitly gauge-invariant formalism and
appears there as an ``intrinsic Coulomb phase'' \cite{JS90} stemming
from the long-range interactions of a colored quark, created initially
at the ``point'' $-\infty^+$ together with its oppositely color-charged
counterpart, then travelling along the plus light-cone direction to the
origin, where it is affected by a hard collision with the photon, thus
changing its route and going along the minus direction to $+\infty^-$.
From this point of view, the soft counter term can be treated as that
part of the TMD PDF which accumulates the residual effects of the
primordial separation of two oppositely color-charged particles,
created at light-cone infinity and being unrelated to the existence of
external color sources.

\paragraph{Conclusions}

The anomalous dimension of the TMD PDF in the {light-cone gauge}
was calculated in the 1-loop order.
It was shown explicitly, how the transverse semi-infinite gauge link
eliminates the dependence from the different pole prescriptions
in the gluon light-cone-propagator.
An anomalous dimension sum rule (ADSR) was introduced, which
allows to study the possible structure of gauge links in the TMD PDF
on the basis of their UV renormalization properties, starting from the
smooth connector which provides the simplest way of gauge-invariance
restoration and obeys simple and well-known RG properties.
A generalized renormalization procedure of the TMD PDF's
was proposed, based on the renormalization of Wilson
exponentials with cusped gauge contours.

\paragraph{Acknowledgements}

I.O.C. is supported by the Alexander von Humboldt Foundation.
This work was supported in part by the Deutsche Forschungsgemeinschaft
under grant 436 RUS 113/881/0, the Russian Federation President's
grant 1450-2003-2, and the Heisenberg--Landau Programme 2008.

\end{document}